\documentclass{ws-p8-50x6-00}

\begin{document}

\title{Mirror Neutrinos and the Early Universe}

\author{Raymond R Volkas}

\address{School of Physics, Research Centre for High Energy Physics,
The University of Melbourne, Victoria 3010, Australia}


\maketitle

\abstracts{I review the construction of the Exact Parity or
Mirror Matter Model and explain how it solves the solar and
atmospheric neutrino problems. The oscillation driven relic
neutrino asymmetry amplification phenomenon is then used to
demonstrate the consistency of the model
with Big Bang Nucleosynthesis. 
}

\section{Lorentz Group: Full and Exact}

The Exact Parity Model (EPM) sees the ordinary particle sector reflected, darkly, in a 
mirror sector.$^{1,2,3}$ It is a phenomenologically acceptable extension of the Standard
Model (SM) of particle physics which displays invariance under Improper Lorentz
Transformations (Parity and Time-Reversal invariance). Remarkably, the invariance or
otherwise of microphysical laws and the physical vacuum under the full Lorentz Group is
still an open question, despite the $V-A$ character of weak interactions.

Almost as a byproduct, the EPM furnishes a unified solution to the solar and atmospheric
neutrino problems; it can also easily accomodate the LSND result.$^3$ In this talk I will
review the resolution of the neutrino anomalies within the EPM. I will also discuss
how the oscillation driven relic neutrino asymmetry amplification mechanism$^4$ ensures
consistency with Big Bang Nucleosynthesis (BBN).$^5$

Consider your favourite parity-violating Lagrangian ${\cal L}(\psi)$ which is invariant
under gauge group $G$. Here I take
${\cal L}$ to be the Lagrangian of the minimal SM augmented by right-handed neutrinos and
nonzero neutrino masses. I also go to the region of parameter space where the see-saw
mechanism operates, so the ordinary left-handed neutrinos are naturally very light
Majorana particles. For every ordinary field $\psi$, introduce a mirror or parity partner
$\psi'$. All ordinary fields are parity-doubled, including gauge and Higgs bosons. The new
Lagrangian
\begin{equation}
{\cal L}(\psi, \psi') = {\cal L}(\psi) + {\cal L}(\psi')
\end{equation}
is a parity-invariant extension of ${\cal L}$ with gauge group $G \otimes G$. The ordinary
and mirror sectors couple by gravitation only at this stage. It is clear, therefore, that
the parity-invariant theory has the same particle phenomenology as the original. It is also
immediately obvious that the mirror sector is astrophysically and cosmologically dark. At
least part of the ``missing mass'' in the universe may be in the form of mirror gas, mirror
stars and the like. It is important to realise that the physical equivalence of the
ordinary and mirror sectors at the microscopic level does not inevitably imply equivalent
macrophysics. Some of the dark matter can be mirror matter {\it without} requiring the
universe to at all stages consist of an equal mixture of ordinary and mirror particles. For
specific mechanisms see Ref.$^6$.

In general, the ordinary and mirror sectors will also interact non-gravitationally. The
interaction Lagrangian ${\cal L}_{\rm int}(\psi,\psi')$ is the sum of all
renormalisable, gauge and parity-invariant terms which couple the $\psi$ to the $\psi'$. In
the EPM based on the SM, these terms are proportional to$^3$
\begin{equation}
F^{\mu\nu}F'_{\mu\nu},\quad \phi^{\dagger}\phi\phi'^{\dagger}\phi',\quad \overline{\ell}_L
\phi (\nu'_L)^c + \overline{\ell}'_R \phi' (\nu_R)^c, \quad \overline{\nu}_R \nu'_L +
\overline{\nu}'_L \nu_R,
\label{Lint}
\end{equation}
where $F^{\mu\nu}$ is the field strength tensor of the weak hypercharge gauge field, $\phi$
is the Higgs doublet, $\ell_L$ is a left-handed lepton doublet while $\nu_R$ is a
right-handed
neutrino. The primed fields are mirror partners. Thus $\nu'_L$ is the mirror partner of
$\nu_R$, and they are both gauge singlets. The leptonic terms have suppressed family 
indices. Each term in Eq.(\ref{Lint}) is multiplied by an {\it a priori} arbitrary
parameter. The full Lagrangian of the EPM is 
\begin{equation}
{\cal L}(\psi,\psi') = {\cal L}(\psi) + {\cal L}(\psi') + {\cal L}_{\rm int}(\psi,\psi').
\label{Lfull}
\end{equation}
The parameters controlling the strength of the first two terms in Eq.(\ref{Lint}), which
induce photon--mirror-photon and Higgs--mirror-Higgs mixing respectively, are constrained
to be small by BBN.

The construction above ensures invariance under the non-standard parity transformation
$\psi \stackrel{P'}{\leftrightarrow} \psi'$. By decomposing the standard $CPT$ operator via
$CPT \equiv P'T'$, we see that non-standard time reversal under $T'$ also follows. The full
Lorentz Group is a symmetry of Eq.(\ref{Lfull}). It is straightforward to check that a
large region of Higgs potential parameter space admits a $P'$ symmetric vacuum.$^2$ For
your 
amusement, observe that the Exact Parity construction is quite similar to supersymmetry in
that both involve extensions of the Proper Lorentz Group and both require particle
doubling. These ``orthogonal'' possibilities are summarised in Fig.1.

\begin{figure}[t]
\epsfxsize=8cm 
\epsfbox{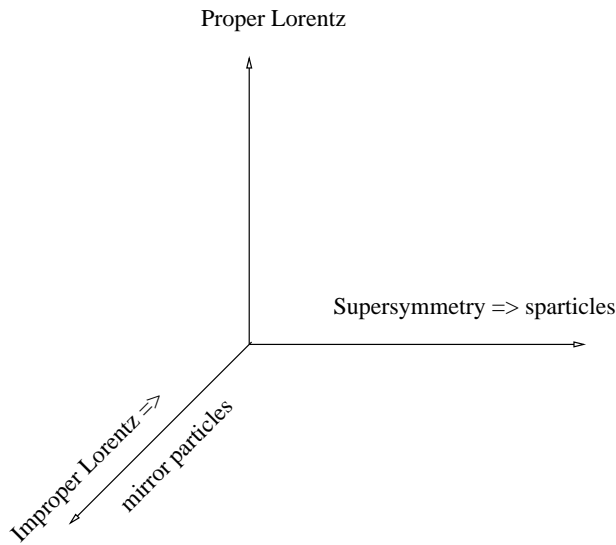} 
\caption{The Exact Parity construction and supersymmetry are similar
in that they both extend the Proper Lorentz Groups and both
require particle doubling.  \label{fig1}}
\end{figure}

\section{Mirror Neutrino Solution to the Solar and Atmospheric
Neutrino Problems}

The last two terms in Eq.(\ref{Lint}) cause ordinary and mirror neutrinos to mix. For full
details about the mass matrix, its diagonalisation and the see-saw mechanism see Ref.$^3$.
It will suffice here to present a fairly model-independent discussion. Suppose some
mechanism, for instance the see-saw as considered above, produces three light ordinary
neutrinos and three light mirror neutrinos. Mixing will also be present. In the region of
parameter space where interfamily mixing is small, three pairs of
parity-eigenstate and therefore maximally-mixed
ordinary and mirror neutrinos are produced. The pairwise maximal mixing is enforced by the
unbroken parity symmetry. The mass eigenstate neutrinos of family $\alpha$, where
$\alpha = e,\mu,\tau$, are given by
\begin{equation}
|\nu_{\alpha \pm}\rangle = \frac{|\nu_{\alpha}\rangle \pm |\nu'_{\alpha}\rangle}{\sqrt{2}}.
\end{equation}
This mass eigenstate pattern is identical to the pseudo-Dirac case as far as terrestrial
experiments are concerned, because the mirror neutrinos are effectively sterile. The EPM is
in part an explicit theory of light sterile neutrinos, with the important additional
feature of pairwise maximal mixing with the corresponding ordinary neutrino. The $\Delta
m^2$ values are in general arbitrary parameters.

The atmospheric neutrino data can be explained by $\nu_{\mu} \to \nu'_{\mu}$ oscillation
with $\Delta m^2_{\mu\mu'} = 10^{-3} - 10^{-2}$ eV$^2$. The observed $\pi/4$ mixing angle
is a successful {\it prediction} of the EPM. Most of the solar neutrino data can be
analogously explained by maximal $\nu_e \to \nu'_e$ oscillations, with $\Delta m^2_{ee'} =
4 \times 10^{-10} - 10^{-3}$ eV$^2$ (the Homestake data point is lower than the EPM
expectation). Neutral current measurements will further test these hypotheses or rule them
out. The LSND results can be accomodated by switching on the appropriate amount of small
interfamily mixing.$^3$

\section{Early Universe Cosmology}

Theories with light sterile or mirror neutrinos give rise to interesting early universe
cosmology. It has been shown that ordinary-sterile (or ordinary-mirror)
neutrino oscillations can dynamically amplify the CP asymmetry of the primordial plasma
through the production of large relic neutrino-antineutrino asymmetries or chemical
potentials.$^{4,5}$ The $\alpha$-like asymmetry $L_{\nu_{\alpha}}$ is defined by
\begin{equation}
L_{\nu_{\alpha}} = \frac{n_{\nu_{\alpha}} - n_{\overline{\nu}_{\alpha}}}{n_{\gamma}},
\end{equation}
where $n_f$ is the number density of species $f$. The oscillation-driven
amplification mechanism can generate asymmetries as high as about $3/8$. This should be
compared to the known baryon asymmetry which is at the $10^{-10}$ level.

A $\nu_{\alpha} \leftrightarrow \nu'_{\beta}$ ordinary-mirror mode
will generate a large asymmetry
prior to the BBN epoch provided the oscillation parameters satisfy$^5$
\begin{eqnarray}
& 10^{-10} \stackrel{<}{\sim} \sin^2 2\theta_{\alpha\beta'} \stackrel{<}{\sim} {\rm few}
\times 10^{-4} \left( \frac{{\rm eV}^2}{|\Delta m^2_{\alpha\beta'}|}
\right)^{\frac{1}{2}},&
\nonumber\\
& \quad \Delta m^2_{\alpha\beta'} < 0,\quad |\Delta m^2_{\alpha\beta'}| \stackrel{>}{\sim}
10^{-4}\ {\rm eV}^2.&
\end{eqnarray}
The small vacuum mixing angle required means that only interfamily $\nu_{\alpha} \to
\nu'_{\beta}$ $(\alpha \neq \beta)$ modes can drive this phenomenon. 

Large neutrino asymmetries have two important consequences. First, they suppress
ordinary-mirror oscillations and hence mirror neutrino production in the early
universe.
Such a suppression mechanism is very welcome, because successful BBN will not be achieved
if the mirror sector is in thermal equilibrium with the ordinary sector during the BBN
epoch. The doubling in the expansion rate of the universe that would result is not
compatible with light element abundance observations. Second, a $\nu_e - \overline{\nu}_e$
asymmetry of the correct magnitude will appreciably affect the neutron-proton
interconversion rates and hence also the primoridal Helium abundance. This can be
conveniently quantified by quoting an {\it effective} number of relativistic neutrino
flavours $N_{\nu,{\rm eff}}$ during BBN. Depending on the sign, an $e$-like asymmetry can
increase or decrease $N_{\nu,{\rm eff}}$ from its canonical value of 3.

A full analysis within the EPM of neutrino asymmetry generation and implications for BBN
can be found in Ref.$^5$. I can provide only a very brief summary here.

The mirror neutrino suppression mechanism relies on the effective potential for the
$\nu_{\alpha} \leftrightarrow \nu'_{\beta}$ in primordial plasma. For temperatures
between the muon annihilation epoch and BBN it is given by
\begin{equation}
V_{\rm eff} = \sqrt{2} G_F n_{\gamma} \left[ L^{(\alpha)} - L'^{(\beta)} - A_{\alpha}
\frac{T^2}{M_W^2} \frac{p}{\langle p \rangle} \right],
\end{equation}
where $G_F$ is the Fermi constant, $M_W$ is the $W$-boson mass and $A_{\alpha}$ is a
numerical factor, $p$ is the neutrino momentum or energy with $\langle p \rangle \simeq
3.15 T$ being its thermal average. The effective asymmetry $L^{(\alpha)}$ is given by
\begin{equation}
L^{(\alpha)} = L_{\nu_{\alpha}} + L_{\nu_e} + L_{\nu_{\mu}} + L_{\nu_{\tau}} + \eta
\end{equation}
where $\eta$ is a small term due to the baryon and electron asymmetries. If the effective
asymmetry is large, then the large matter potential will suppress the associated
oscillation mode. Under the conditions discussed above, a small angle $\nu_{\alpha}
\leftrightarrow \nu'_{\beta}$ $(\alpha \neq \beta)$ mode will generate a brief period of
exponential growth of $L_{\nu_{\alpha}}$ at a critical temperature $T_c$ given
approximately by $T_c \sim 16[(|\Delta m^2_{\alpha\beta'}|/{\rm eV}^2) \cos
2\theta_{\alpha\beta'}]^{1/6}$ MeV.

Consider the parameter space region
\begin{equation}
m_{\nu_{\tau\pm}} \gg m_{\nu_{\mu\pm}} \gg m_{\nu_{e\pm}}
\end{equation}
with the $\nu_{\alpha} - \nu'_{\alpha}$ splittings relatively small. Let the interfamily
mixing angles be small. Focus on the
$\tau-\mu$ subsystem, set $\Delta m^2_{\mu\mu'}$ to the atmospheric anomaly range,
and set $\Delta m^2_{\tau\tau'} = 0$ for simplicity. The $\nu_\tau \leftrightarrow
\nu'_{\mu}$ mode satisfies the conditions for generating a large $L_{\nu_{\tau}}$.
Considered by itself, this generates a large $\mu\mu'$-like effective asymmetry 
$L^{(\mu)} - L'^{(\mu)}$ and so 
suppresses the maximal $\nu_{\mu} \leftrightarrow \nu'_{\mu}$ mode. This prevents
the $\nu'_{\mu}$ from thermally equilibrating. However, it turns out that the
$\nu_{\mu} \leftrightarrow \nu'_{\mu}$ tries to destroy its effective $\mu\mu'$-like
asymmetry.
For a given $\Delta m^2_{\mu\mu'}$, there is a region of $(\Delta m^2_{\tau\mu'}, \sin^2
2\theta_{\tau\mu'})$ parameter space in which $L^{(\mu)} - L'^{(\mu)}$ is not efficiently
destroyed, and
another region in which it is. To analyse this complicated dynamics, one must solve a set
of coupled Quantum Kinetic Equations. The outcome is displayed in Fig.2. The three solid
lines correspond to $\Delta m^2_{\mu\mu'} = 10^{-3},10^{-2.5},10^{-2}$ eV$^2$ from bottom
to top. Above the solid line, $\nu'_\mu$ production via $\nu_{\mu} \to
\nu'_{\mu}$ is negligible. Below the line, the
$\nu'_{\mu}$ is brought into thermal equilibrium. The dot-dashed line refers $\nu'_{\mu}$
production by the asymmetry-creating mode $\nu_{\tau} \to \nu'_{\mu}$. To the left of the
line, $N_{\nu,{\rm eff}} < 3.6$. The BBN ``bound'' of $3.6$ was chosen for illustrative
purposes only. It is interesting that the $\Delta m^2_{\tau\mu'}$ values required for
consistency with BBN are compatible with a $\nu_\tau$ mass in the hot dark matter
range (shaded band).

\begin{figure}[t]
\epsfxsize=9cm 
\epsfbox{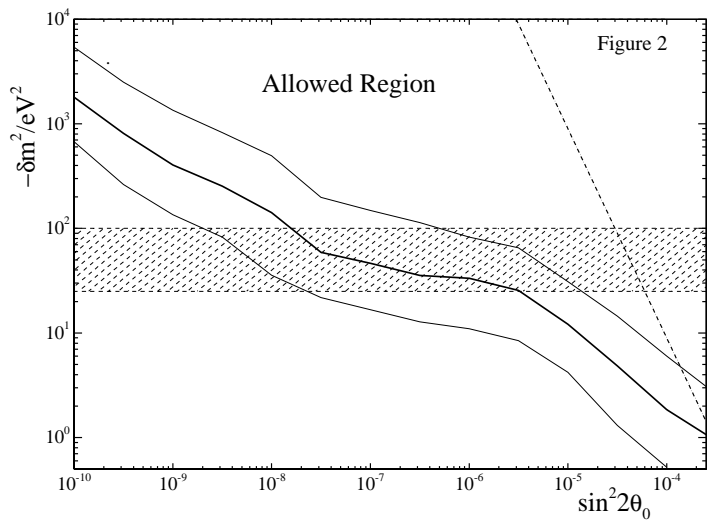} 
\caption{The region of $\nu_\tau \leftrightarrow \nu'_\mu$ oscillation parameter space
which leads to consistency of the $\nu_\mu \to \nu'_\mu$ solution of the atmospheric
neutrino problem with BBN. See text for full discussion. \label{fig2}} 
\end{figure}

A lengthy analysis allows one to also estimate the $e$-like asymmetry that gets generated
if the $e$ family couples non-negligibly with the more massive families.$^5$ It turns out
that
the mass-squared difference between the first and second families, denoted $\Delta m^2_{\rm
small}$ in Ref.$^5$, plays an important role.  The results for $\delta N_{\nu,{\rm eff}}
\equiv N_{\nu,{\rm eff}} - 3$ depend on the sign of the asymmetry generated, which
unfortunately cannot be predicted at present because it depends on the unknown initial
values of the asymmetries.$^4$ The point is simply that an unknown mechanism, not
associated
with light neutrino oscillations, may generate nonzero asymmetries at some much earlier
epoch. While the final magnitudes of the asymmetries around the time of BBN are insensitive
to the initial values provided the latter are small, the same is not true for the overall
signs of the asymmetries. The signs could also be affected by spatial inhomogeneities in
the baryon asymmetry.$^7$ The convenient quantity $\delta N_{\nu,{\rm eff}}$ will be
greater or
less than zero depending on the sign of $L_{\nu_e}$. The results show that $\delta
N_{\nu,{\rm eff}}$ gets to the $\pm 1$ regime for $\Delta m^2_{\rm small}$ values in the
eV$^2$ to few eV$^2$ range. Interestingly, this puts the $\nu_e - \nu_\mu$ mass-squared
difference in the LSND regime. Improved primordial abundance measurements are needed before
definite conclusions can be drawn about the cosmologically favoured first-second family
mass splitting in the EPM.

At this meeting, A. Dolgov$^8$ argued that the oscillation-driven neutrino asymmetry
amplification mechanism cannot generate asymmetries as high as claimed in Refs.$^{4,5}$.
I do not
agree. Space limitations prevent discussion of these issues here. See Ref.$^9$ for comments
on the criticisms of Dolgov et al.

\section{Conclusions}

The Mirror or Exact Parity Model solves the solar and atmospheric neutrino problems in a
cosmologically consistent way. It is also compatible with the LSND result. The role of
neutrino oscillations prior and during BBN is quite remarkable because of the relic
neutrino asymmetry amplification phenomenon. If they exist, light mirror
neutrinos will make life and the universe even more interesting!

\section*{Acknowledgments}
Warmest thanks to Rachel Jeannerot, Goran Senjanovic and Alexei Smirnov
for organising this interesting meeting. I acknowledge lively discussions during
the meeting with Z. Berezhiani, A.
Dolgov, K. Enqvist, K. Kainulainen, S. Pastor and A. Sorri on topics relevant to this
paper.
I would like to thank Nicole Bell, Roland Crocker, Pasquale Di Bari,
Robert Foot, Keith Lee, Paolo Lipari, Maurizio Lusignoli and Yvonne Wong
for their collegiality and their insights. This work was supported
by the Australian Research Council.

\section*{References}

Unfortunately, space does not permit full referencing here. The numbers appearing after
references below correspond to the appropriate citations contained in Ref.$^5$.
[1] 1.\ [2] 3.\ [3] 4.\ [4] 24; 25; 26; 28; 29; P. Di Bari, P. Lipari and M. Lusignoli,
hep-ph/9907548, Int.\ J. Mod.\ Phys.\ (in press); P. Di Bari, these proceedings.\ [5] R.
Foot and R. R. Volkas,
hep-ph/9904336, Phys.\ Rev.\ D (in press).\ [6] E. W. Kolb, D. Seckel and M. S. Turner,
Nature {\bf 514}, 415 (1985); H. M. Hodges, Phys.\ Rev.\ D {\bf 47}, 456 (1993); V.
Berezinsky and A. Vilenkin, hep-ph/9908257; Z. Berezhiani, these proceedings.\ [7] P. Di
Bari, hep-ph/9911214.\ [8] A. Dolgov, these proceedings.\ 
[9] P. Di Bari and R. Foot, hep-ph/9912215; A. Sorri,
hep-ph/9911366.

\end{document}